\documentclass[10pt,english]{article}
\usepackage{subfigure}
\usepackage{graphicx}
\usepackage{float}
\usepackage[T1]{fontenc}
\usepackage[latin9]{inputenc}
\usepackage[margin=1.25in]{geometry}
\usepackage{float}
\usepackage{amsmath}
\usepackage{amsbsy}
\usepackage{setspace}
\usepackage{amssymb}
\usepackage{esint}
\usepackage{cite}
\newfloat{algorithm}{tbp}{loa}
\floatname{algorithm}{Algorithm}
\usepackage{hyperref}
\usepackage{listings}

\newlength\myindent
\makeatletter

\floatstyle{ruled}
\newfloat{algorithm}{tbp}{loa}
\floatname{algorithm}{Algorithm}

\usepackage{babel}

\begin{document}

\title{Reaction Diffusion TAP}

\author{M. Andrecut}

\date{September 5, 2023}

\maketitle
{

\centering Unlimited Analytics Inc.

\centering Calgary, Alberta, Canada

\centering mircea.andrecut@gmail.com

} 
\bigskip 
\begin{abstract}

The recently introduced Theory of the Adjacent Possible (TAP) is a model of combinatorial innovation 
aiming to explain the "hockey-stick" upward trend of human technological evolution, where an explosion 
in the number of produced items with increasing complexity suddenly occurs. In addition, the TAP model was also 
used to explain the rapidly emerging biological complexity. Inspired by TAP here we propose a reaction-diffusion 
system aiming to extend the model in both space and time. We show that the new model exhibits similar 
characteristics to the TAP model, like the sudden increase in the production of items, after a longer period of 
slow growth. The new model also exhibits wave propagation of "innovation", resulting in 
self-sustained complex interference patterns. 

\smallskip 


\end{abstract}

\section{Introduction}

After a long period ($\sim10^5$ years) of slow human "innovation", the industrial revolution started around the 19 century, 
generating a sudden increase of goods, accompanied also by a similar increase in population. This sudden explosion was 
metaphorically associated with the "hockey stick" upward trend of economic growth, which recently has produced a larger number  
of items of increasing complexity. The Theory of the Adjacent Possible (TAP) was initially introduced as a model of 
technological change aiming to explain this combinatorial explosion of the industrial revolution \cite{key-1}, \cite{key-2}. 
The main idea of TAP is that "what exists now does not necessarily cause, but certainly enables what could arise next" \cite{key-3}. 
Which means that the complexity of the newly-created "items" arises from the combination of features of already existing "items", 
which in turn have been created by combining features from earlier items, and so on. Assuming that $M_t$ is the number of items in the 
economy at time $t$, then the TAP equation can be written as:
\begin{equation}
M_{t+1} = M_t + \sum_{i=1}^{M_t} \alpha_i \binom{M_t}{i},
\end{equation}
where $0< \alpha_i<1$ is a decreasing sequence $\alpha_{i}>\alpha_{i+1}$, accounting for the decreasing ability of producing new 
useful combinations among an increasing number of items. The self-feeding (positive feedback) iterative and combinatorial mechanisms 
embedded in this equation lead to an explosion of goods created in finite time \cite{key-4}. 

The TAP model was also used to explain the rapidly emerging biological complexity, where the produced "items" could be the molecular species 
present in the biosphere \cite{key-5}, \cite{key-6}. These molecules interact via (spontaneous) chemical reactions between arbitrary combinations of currently existing molecules, 
and produce new molecular species. In this case the parameters $\alpha_i$  may me interpreted as reaction propensities. The biosphere is therefore 
also characterized by such an explosive growth in the number of molecular species. 

In this article we propose a reaction-diffusion system aiming to extend the TAP model in both space and time. 
We show that the new model exhibits the explosive growth characteristic of the TAP model, and in addition also exhibits wave propagation of "innovation", 
resulting in complex interference patterns, able to locally produce large fluctuations of "innovation" values, that can sustain the production of new "items", or eventually lead to extinction.

\section{Reaction diffusion model}

The original TAP equation (1) has several shortcomings. The main variable $M_t$ is supposed to be an integer, yet the other parameters are real numbers, 
which means that after a single iteration $M_t$ is not an integer anymore, and this leads to a difficult interpretation of the results. 
Also, $M_t$ increases very fast and the sum cannot be computed anymore, since it quickly results in very large numbers, and even the authors are truncating the sum at $i_{max}=4$ 
in their numerical estimations \cite{key-3}-\cite{key-6}.
In order to avoid these inherent computational difficulties, our approach here is different, and it is based on the more traditional population growth models \cite{key-7}. 

We denote by $x(t)$ the concentration of "items" existing at time $t$. Also, we assume that the "universe" has a finite carrying capacity $k>0$, and therefore 
$x(t)$ represents a fraction of this capacity. Moreover, we assume that at each time $t$ the derivative of $x(t)$ is given by a "logistic like" function:
\begin{equation}
\frac{dx}{dt} = ax(1-k^{-1}x)f(x), 
\end{equation}
where $f(x)$ is a function capturing the combinatorial "innovation" required to produce new "items". 
The other factors correspond to the traditional logistic model with a growth constant $a>0$. 
In order to simplify the notation and the presentation it is convenient to assume that the carrying capacity is set to  $k=1$. 

The solution of the standard logistic model: 
\begin{equation}
\frac{dx(t)}{dt} = ax(t)(1-x(t)), \quad x(0) = x_0,
\end{equation}
is given by:
\begin{equation}
x(t) = \frac{x_0}{x_0+(1 - x_0)e^{-at}}.
\end{equation}
One can see that the growth saturates and it is limited by the carrying capacity:
\begin{equation}
\lim_{t\rightarrow\infty} x(t) = \lim_{t\rightarrow\infty} \frac{x_0}{x_0+(1 - x_0)e^{-at}} = 1.
\end{equation}

The equilibrium condition (zero growth) for the logistic model is:
\begin{equation}
\frac{dx}{dt} = ax(1-x)=0.
\end{equation}
This equation has two solutions: $x=0$ and $x=1$, which can be used to characterize the stability of the logistic model. 

The growth function of the logistic model is:
\begin{equation}
h(x)=ax(1-x),
\end{equation}
and its derivative with respect to $x$ is:
\begin{equation}
\frac{dh}{dx} = a(1-2x).
\end{equation}
The values of the derivative at the solution points are:
\begin{equation}
\frac{dh}{dx} \bigg\vert_{x\rightarrow0} = a > 0, \quad \frac{dh}{dx}\bigg\vert_{x\rightarrow1}  = -a < 0,
\end{equation}
which means that $x=0$ is an unstable solution, while $x=1$ is a stable solution, attracting the dynamical evolution of the "items" population. 
Also, for $x=1/2$ the model exhibits a maximum growth $h^*(1/2)=a/4$. 

In our approach we assume that the combinatorial "innovation" function is given by:
\begin{equation}
f(x) = \sum_{i=0}^n \binom{n}{i} \alpha^ix^i - 1 = (1+\alpha x)^n-1, \quad n>0.
\end{equation}
This means that at each moment in time 
all the combinations up to $n$ are taken into account, but they are discounted by the decreasing parameters $\alpha_i = \alpha^i$, where $0<\alpha<1$. 
Here we have removed $1$ from the sum in order to avoid constant growth. Also, it is worth noting that the quantity:
\begin{equation}
xf(x)= \sum_{i=0}^n \binom{n}{i} \alpha^ix^{i+1} - x= \sum_{i=1}^n \binom{n}{i} \alpha^ix^{i+1},
\end{equation}
contains only "combinatorial" terms: $x^2,x^3,...,x^{n+1}$. For example $x^2$ corresponds to binary combinations, while $x^3$ corresponds to ternary combinations etc. 
Thus the equation contains all the desired combinatorial terms. 

We should also note that $f(x) > 0$ for all $n>0$ and $0<x<1$, and therefore the stability conditions for the TAP differential model are the same as for the standard logistic model, 
and as a consequence $x=1$ is still a stable solution attracting the dynamics of the produced "items". 

\begin{figure}[!ht]
\centering \includegraphics[width=9.5cm]{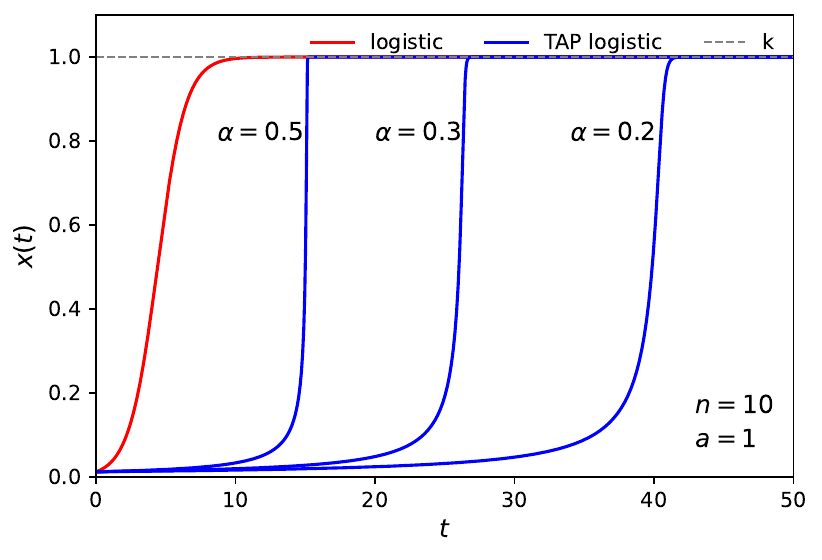}
\centering \includegraphics[width=9.5cm]{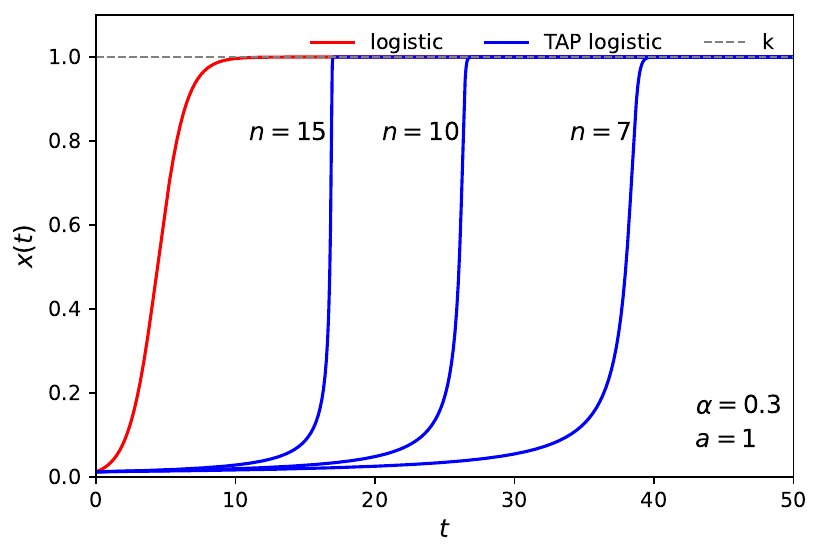}
\centering \includegraphics[width=9.5cm]{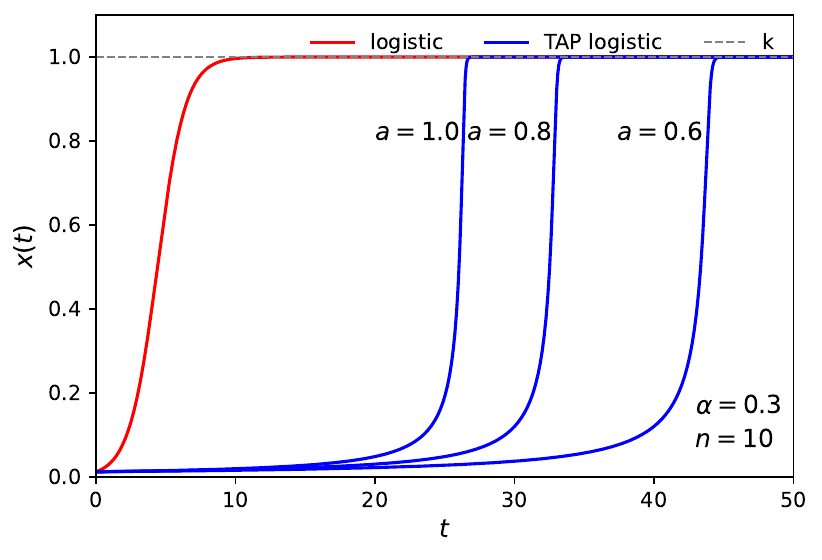}
\caption{The difference between the standard logistic, and TAP logistic growth models, and the role of the $\alpha$, $n$ and $a$ parameters. }
\end{figure}

The maximum growth of the differential TAP model is obtained at $x=1/2$, and besides the constant growth $a$ it also depends on the "innovation" function parameters ($\alpha, n$):
\begin{equation}
h_{TAP}^*(1/2) = \frac{a}{4} \left[ \left( 1 + \frac{\alpha}{2} \right)^n -1 \right] .
\end{equation}

Therefore the differential TAP logistic equation can be written as following:
\begin{equation}
\frac{dx}{dt} = ax(1-x)[(1+\alpha x)^n-1], \quad x(0) = x_0,
\end{equation}
where $a>0$, $0<\alpha <1$, and $n>0$. 

In Figure 1 we show the difference between the TAP logistic equation and the traditional logistic equation, and the role of the 
$\alpha$, $n$ and $a$ parameters. The concentration of "items" at the initial condition $t=0$ in all cases was set to $x_0=0.012345$. 
In the first case (Fig. 1, top) we kept the parameters $n=10$ and $a=1.0$ constant, and we varied $\alpha \in \{0.2,0.3,0.5\}$. 
In the second case (Fig. 1, middle) we kept $\alpha=0.3$ and $a=1.0$ constant, and we varied $n \in \{7,10,15 \}$. In the third case (Fig. 3, bottom), we kept $\alpha=0.3$ and $n=10$ constant, 
and we varied $a \in \{0.6,0.8,1.0\}$. 
One can see that in all three cases the TAP logistic equation shows a longer and much slower increase than the standard logistic equation, 
followed by a sudden explosive increase, which is similar to the behavior of the original TAP equation. All three parameters play a significant role in 
controlling the delay of the explosive behavior. A combination of higher values of $n$ and $\alpha$, with a relatively lower value of the growth constant $a$ is 
the main characteristic of the TAP model behavior.

Another shortcoming of the original TAP model (1) is that it assumes unlimited resources to perform the "innovation" and production work. In reality, the "innovation" is conditioned by 
the existence of (human) capital and other strictly necessary resources (like energy for example). In order to model these resources we introduce a second variable $y$, the "resources", 
which is governed by a standard logistic growth. 
We also assume that the production of "items" is conditioned by how many "resources" are available to sustain the innovation work (or the reactions in the biosphere), and also 
that both the "items" and the "resources" are removed with the rate $c_x>0$, and respectively $c_y>0$ from the system, due to natural degradation or death. 
Therefore, we can write the following TAP system of coupled differential equations:
\begin{equation}
\frac{dx}{dt} = ax(1-x)f(x)y-c_x x, \quad x(0) = x_0,
\end{equation}
\begin{equation}
\frac{dy}{dt} = by(1-y) - ax(1-x)f(x)y-c_y y, \quad y(0) = y_0.
\end{equation}
In the first equation the number of produced "items" $ax(1-x)f(x)y$ is conditioned by the existence of the "resources" $y$, and subsequently the same quantity is removed from the second equation, since 
the same amount of "resources" have been used. 

Numerical simulations of the above system show the existence of three distinct regimes, depending on the relative values of the model parameters. 
In order to illustrate these distinct regimes we kept the following parameters constant: $a=1$, $\alpha=0.1$, $n=10$, $c_x=0.02$, $c_y=0.01$, and we only varied the "resources" growth 
constant as $b \in \{0.2,0.1,0.075\}$, as shown in Figure 2. 

\begin{figure}[!ht]
\centering \includegraphics[width=8.5cm]{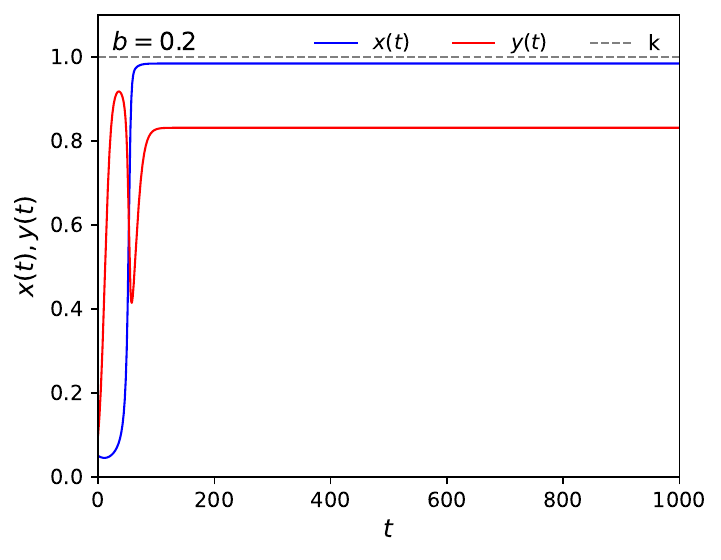}
\centering \includegraphics[width=8.5cm]{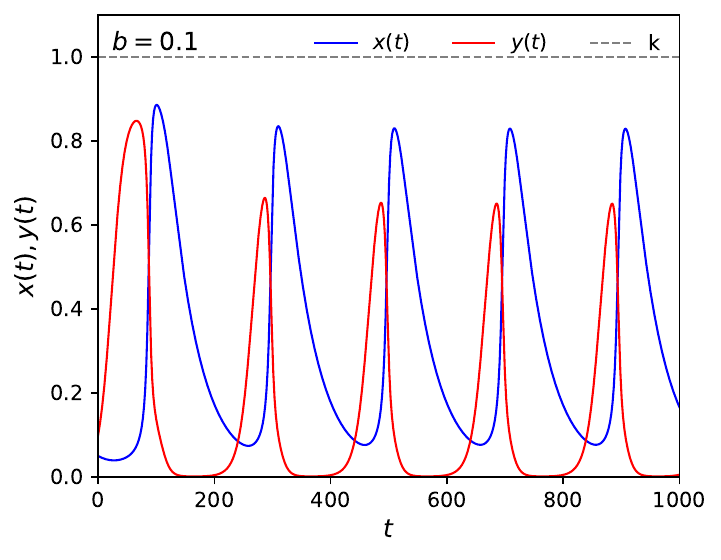}
\centering \includegraphics[width=8.5cm]{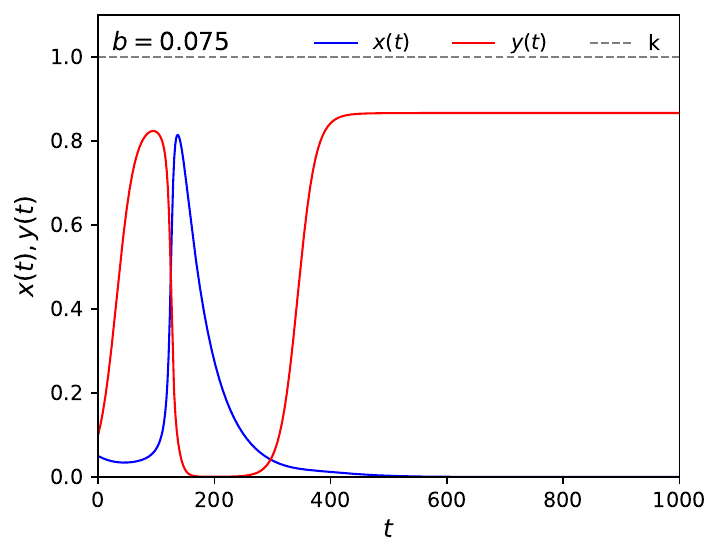}
\caption{Three different dynamical regimes of the differential TAP system.}
\end{figure}

In the first regime the "resources" can grow fast enough ($b=0.2$) in order to sustain a "hockey-stick" type growth (Fig. 2, top). 
The second regime is characterized by sustained oscillations ($b=0.1$), where the fast increase in production of "items" depletes the resources, which leads to a decrease in "items" production, 
which subsequently gives the opportunity for a new increase in "resources", and so on (Fig. 2, middle). In the third regime, 
the "resources" cannot grow fast enough ($b=0.075$) to maintain the "items" production, which collapses after a short transient (Fig. 3, bottom). 
One can see that the behavior of the system resembles the behavior of a predator-prey model \cite{key-7}, where the innovation of "items" (the predator) is chasing the "resources" (the prey).  

The above system of differential equations can be easily extended in space by assuming that both the produced "items" and the "resources" 
can diffuse with the coefficients $d_x>0$ and $d_y>0$. We can therefore write the following system describing the dynamics of the 
reaction diffusion TAP model:
\begin{equation}
\frac{dx}{dt} = d_x\nabla^2x + ax(1-x)f(x)y-c_x x, \quad x(0) = x_0,
\end{equation}
\begin{equation}
\frac{dy}{dt} = d_y\nabla^2y + by(1-y) - ax(1-x)f(x)y-c_y y, \quad y(0) = y_0.
\end{equation}

Let us now consider a couple of relevant numerical examples. 
The numerical implementation is based on the finite difference method using periodic (toroidal) boundary conditions. The following parameters are kept constant 
during the simulations: $d_x=d_y=2^{-7}$, $c_x=0.02$, $c_y=0.01, a = 1, b=0.1, \alpha=0.1$. 
The simulations are performed on a grid of $256\times256$ cells, and the "resources" $y$ are initially randomly distributed on the grid 
with values $0<y<1$. The Python code for simulating and visualizing the reaction diffusion TAP system is given in the Appendix.

\subsection{Localized initial condition}

Let us first assume that the initial "items" are distributed homogeneously in a small square ($20\times20$) in the center of the grid, and the initial values on each cell 
of the small square are $x(0)=0.123$, while all the rest are set to zero (Fig. 3, top, $t=0$). The system evolves in time and one can see that due to the growth and diffusion mechanisms the "resources" homogenize 
in space, and the system generates "innovation" waves that propagate by producing "items" and consuming "resources" (Fig. 3, middle, $t\simeq 10^3$ time steps). Once the waves go beyond the grid limits 
they return on the opposite side (simulating a toroid), and interfere with the other waves. This process results in a chaotic mixing of "innovation" and production waves, resulting 
in chaotically distributed regions with high or low concentrations of "items" (Fig. 3, bottom, $t\simeq 10^5$ time steps). This is a typical scenario where a localized "innovation" can spread over the entire finite "world".
\begin{figure}[!ht]
\centering \includegraphics[width=12.5cm]{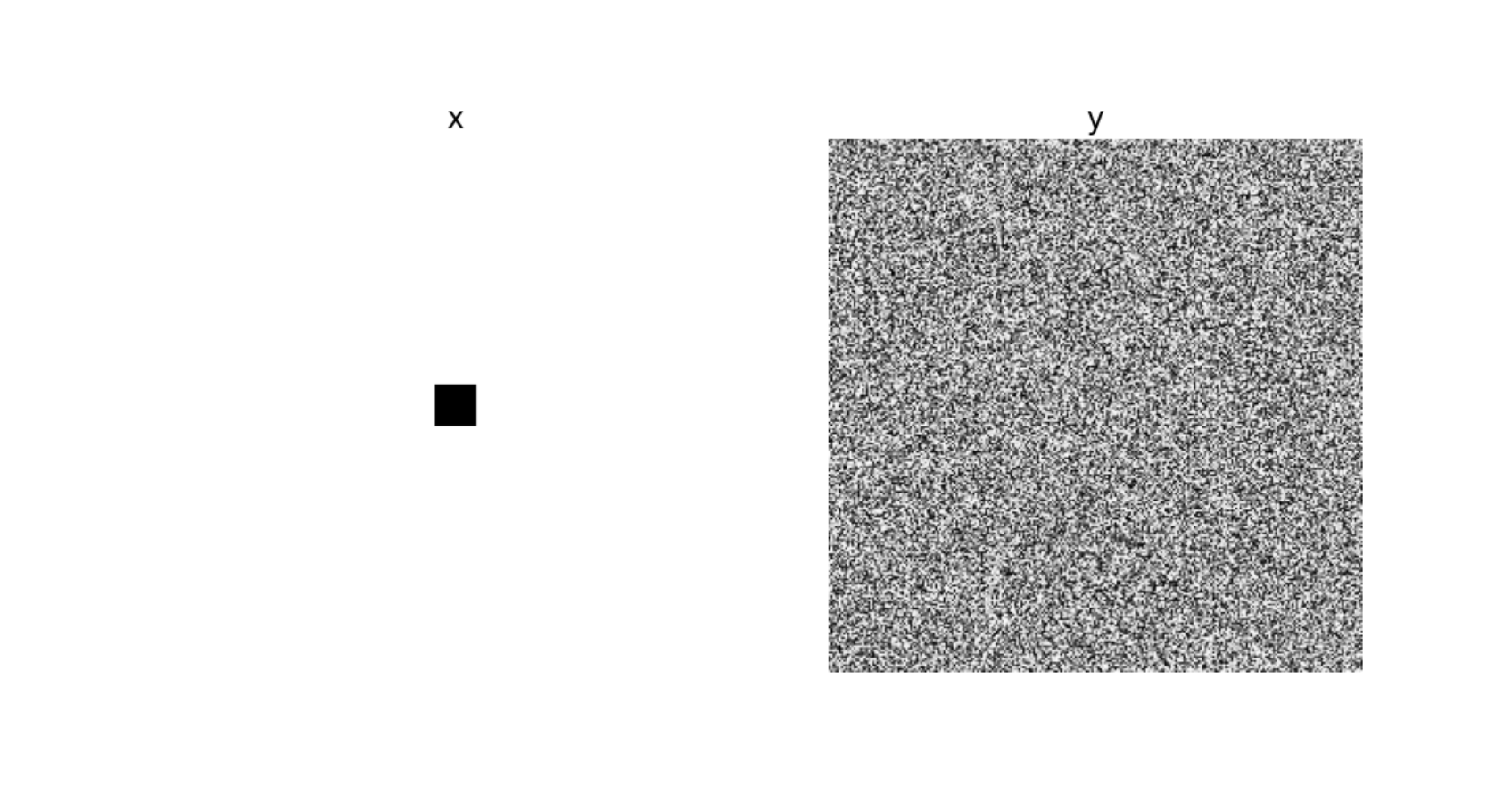}
\centering \includegraphics[width=12.5cm]{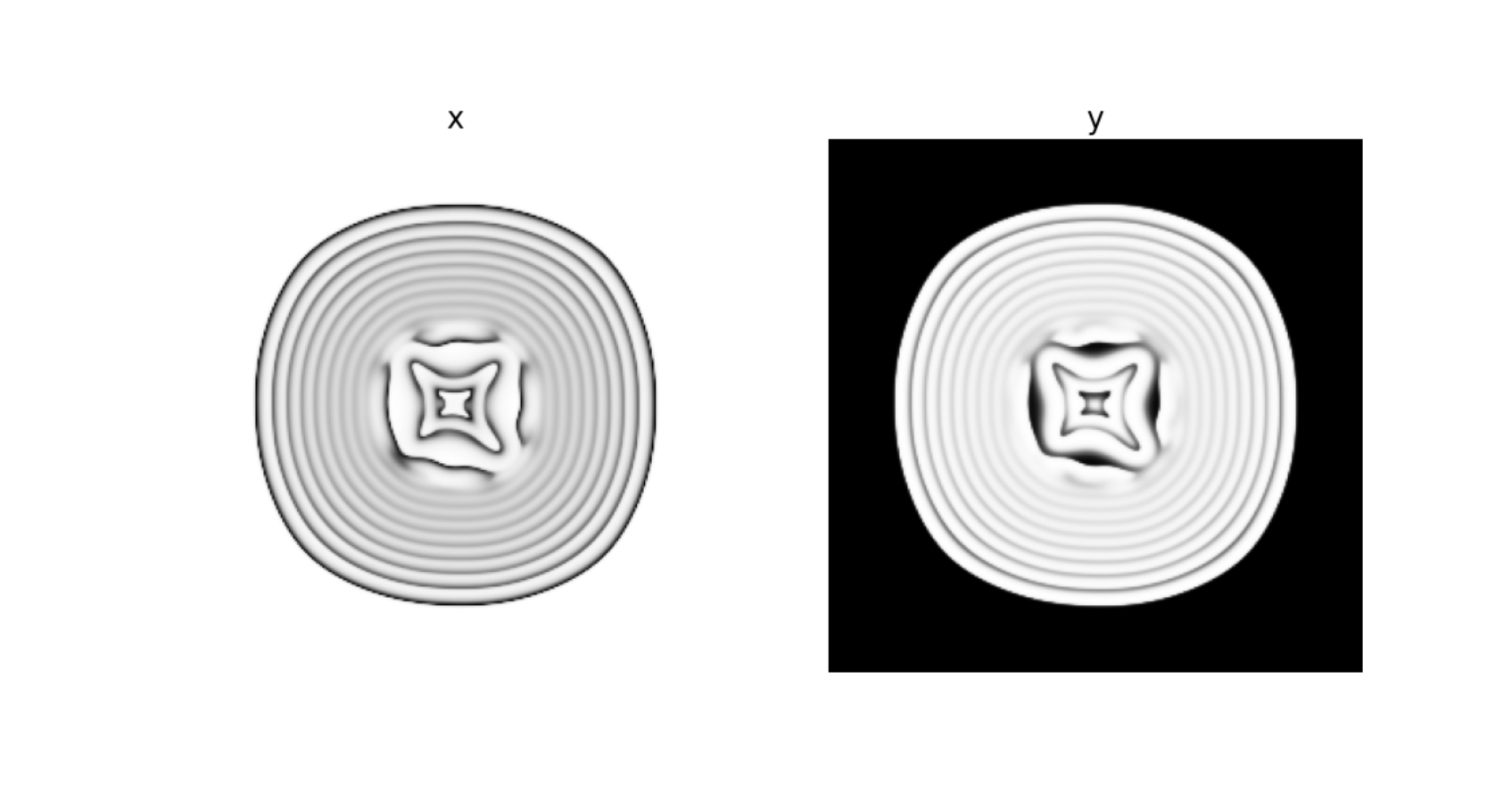}
\centering \includegraphics[width=12.5cm]{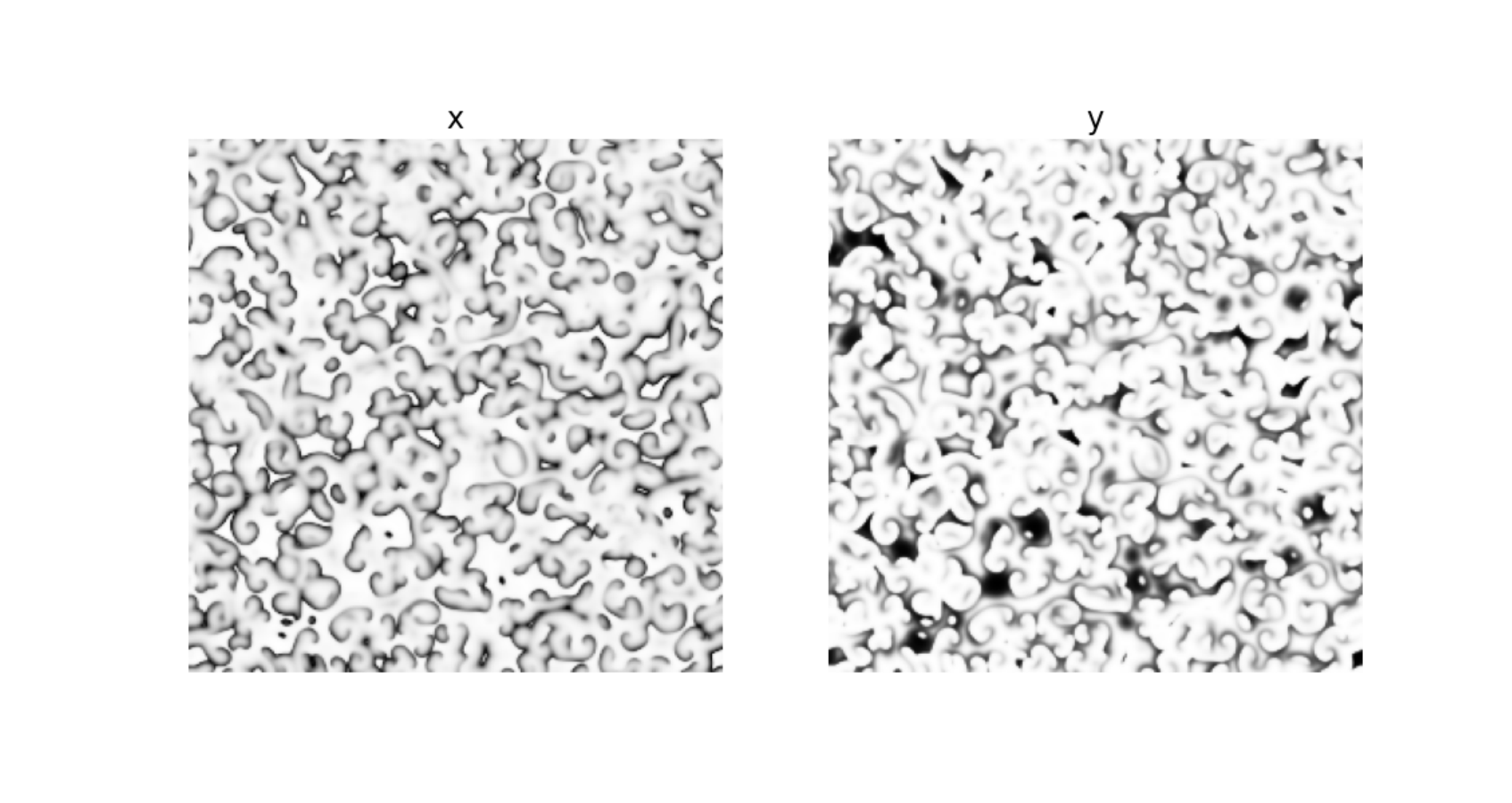}
\caption{Localized initialization (top); Wave propagation (middle); Chaotic mixing phase (bottom).}
\end{figure}

\subsection{Randomly distributed initial condition}

In the second example we assume that the initial "items" are also randomly distributed over the entire grid (Fig. 4, top, $t=0$), 
similar to the initial "resources". After a short transient several "bubbles" of "innovation" 
arise and their waves start to propagate and interfere creating large fluctuations in the production of "items" in the system (Fig. 4, middle, $t \simeq 10^3$). 
After a longer period of time the wave interference process described above creates a similar chaotic mixing observed in the first experiment (Fig. 4, bottom, $t \simeq 10^5$). 
This chaotic mixing is just an effect resulting from the fact that the "world" considered here is finite. 
In an infinite "universe" the "innovation" waves should be able to maintain their deterministic behavior. 
\begin{figure}[!ht]
\centering \includegraphics[width=12.5cm]{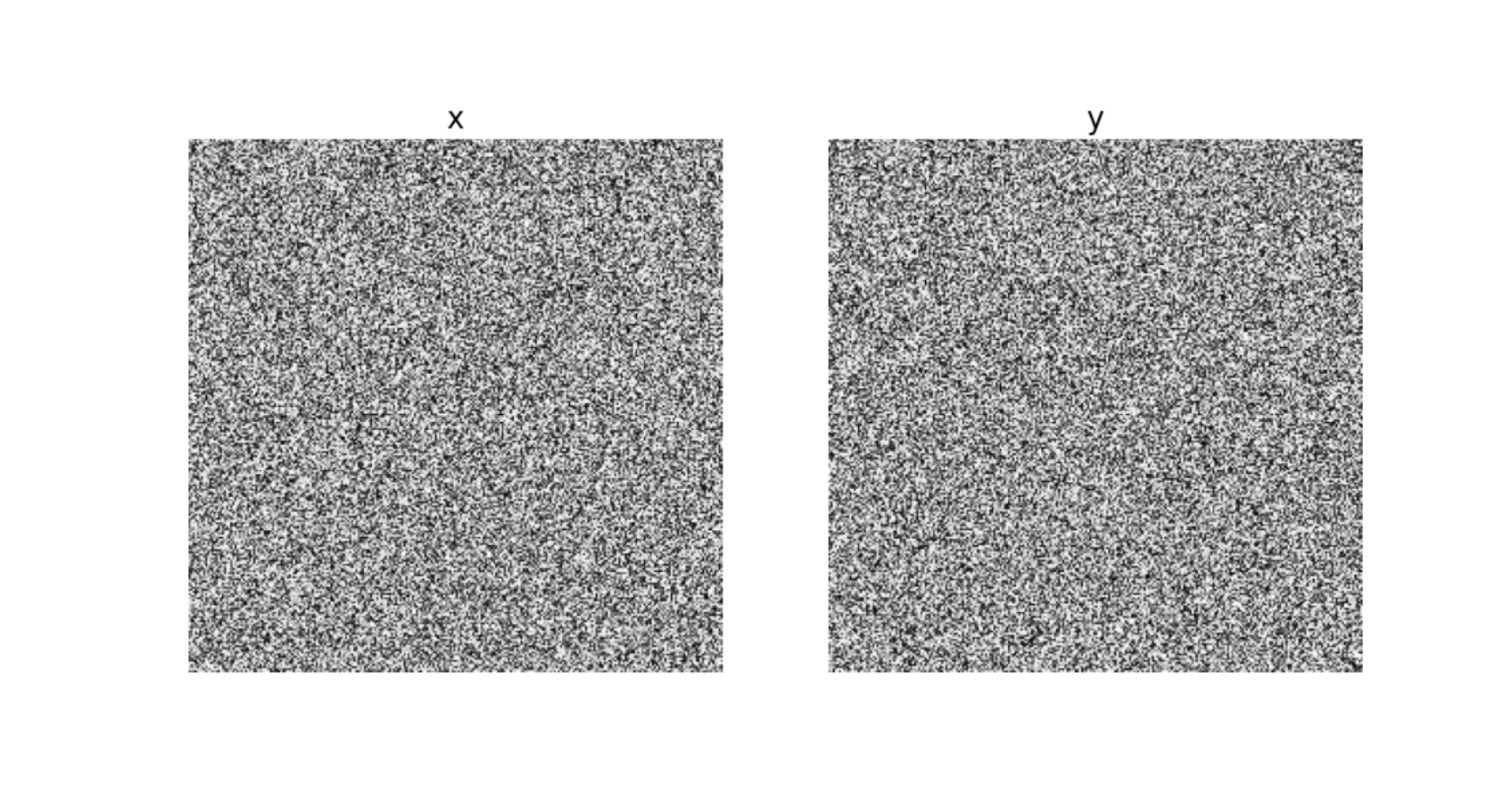}
\centering \includegraphics[width=12.5cm]{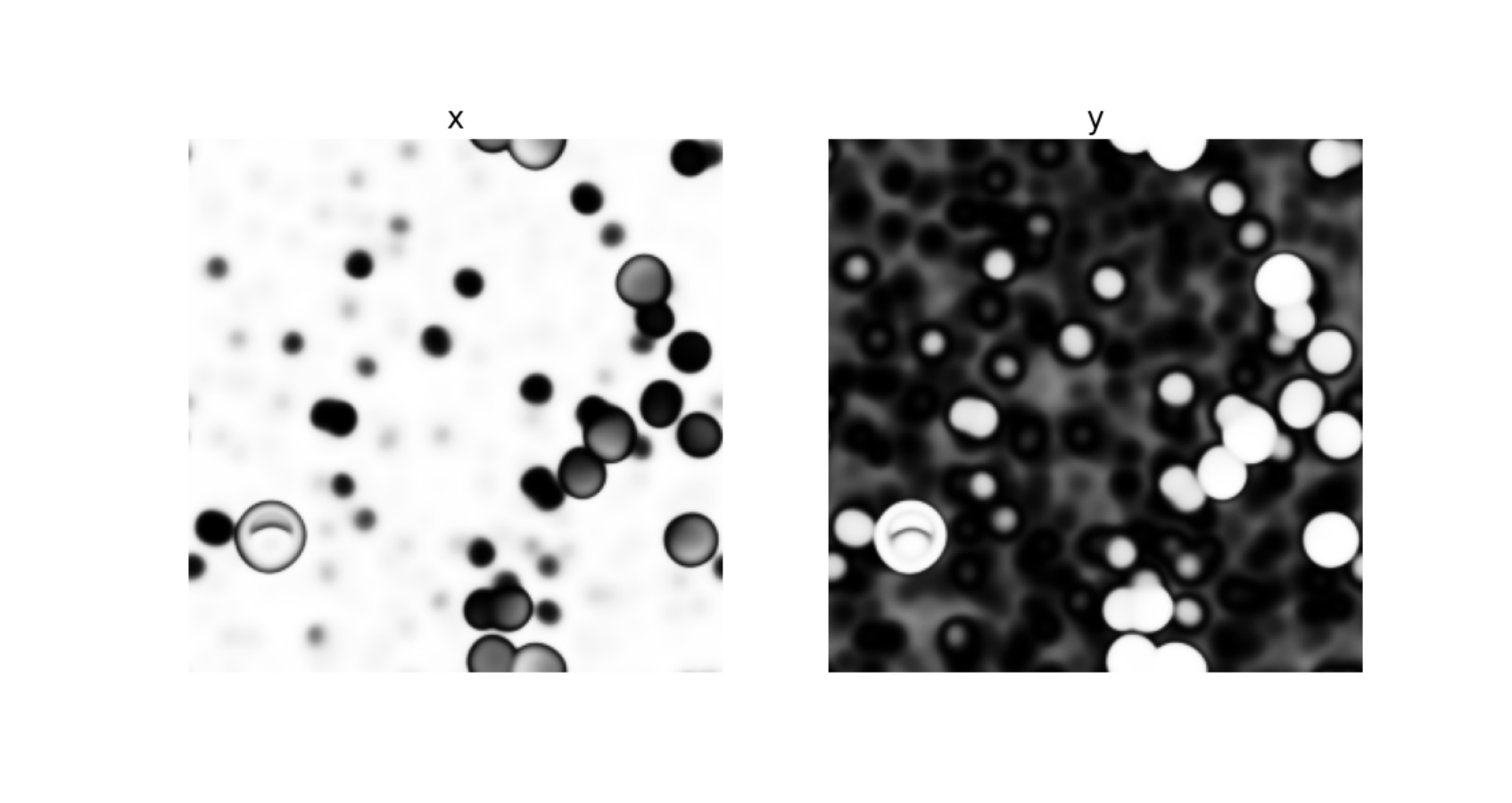}
\centering \includegraphics[width=12.5cm]{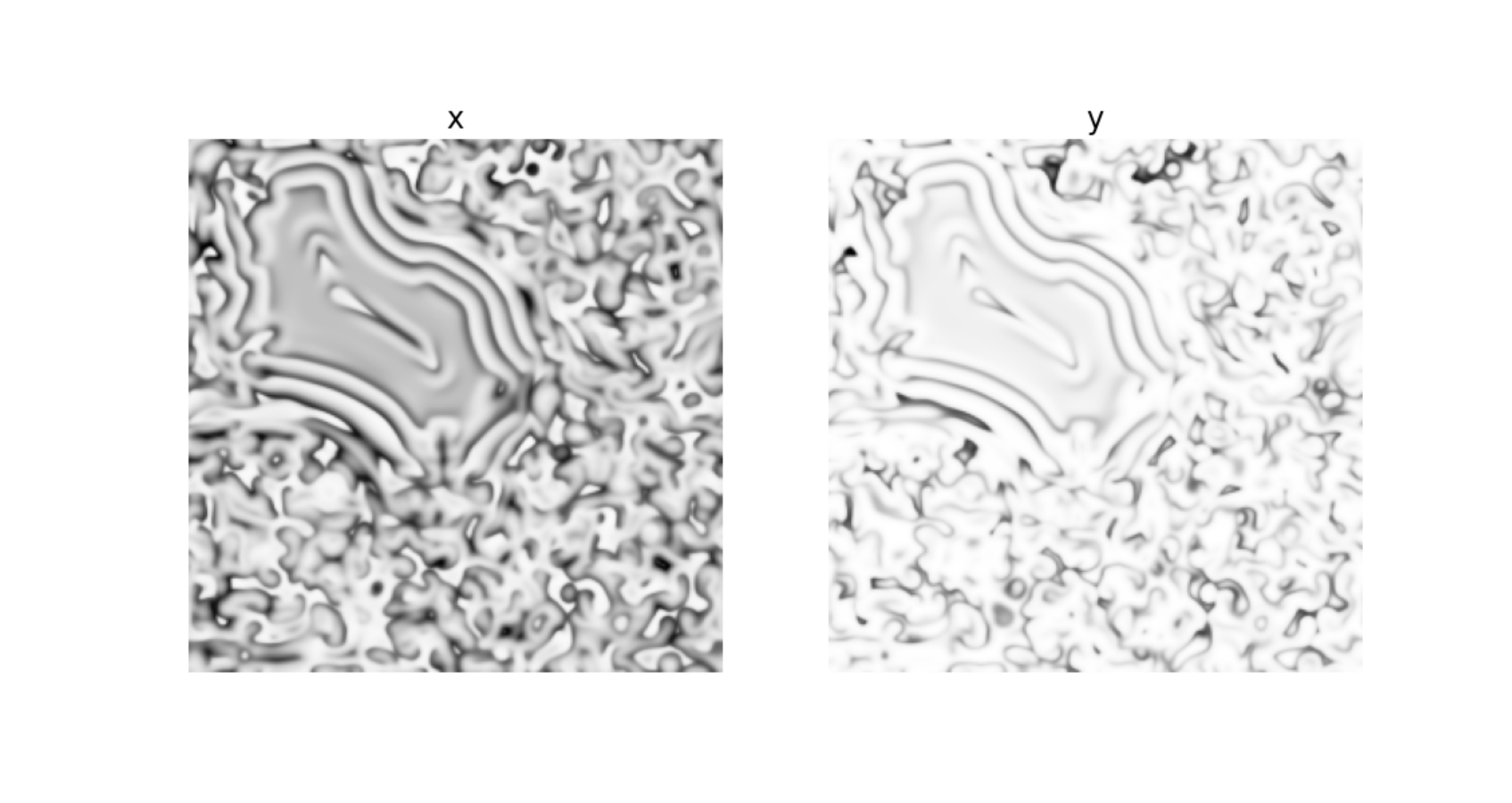}
\caption{Random initialization (top); Wave propagation (middle); Chaotic mixing phase (bottom).}
\end{figure}

\section*{Conclusion}

The TAP model was introduced to explain the "hockey-stick" upward trend observed in the industrial revolution, 
and in the emerging biological complexity of the biosphere. Due to its very strong nonlinearity, the TAP 
model blow-up is sudden, and difficult to predict. The initial growth is slow, but once it approaches a 
critical region it suddenly accelerates and explodes. While this is great for the ability of the system 
to "innovate" and produce a combinatorial number of items, it is also worrisome regarding the ability of the 
system to sustain itself, due to the equally sudden and inevitable depletion of "resources". 
Here we have introduced a reaction-diffusion system that extends the TAP model in both space and time. 
We have shown that the new model exhibits the explosive growth characteristic of the initial TAP model, and in addition also exhibits wave propagation of "innovation" and production, 
resulting in complex interference patterns, able to locally produce high "innovation" values, and to sustain an increased production of new "items".

\section*{Appendix}

The Python code for the reaction diffusion TAP model. 

\begin{footnotesize}
\begin{verbatim}
import numpy as np
import matplotlib.pyplot as pl
import matplotlib.animation as animation
from matplotlib.colors import Normalize

def lap(x,d):
    xx = -4*x.copy()
    for q in [(-1, 0),(0,-1),(0, 1), (1, 0)]:
        xx += np.roll(x, q, (0,1))
    return d*xx
    
def rd(x, y):
    b, d = 0.1, 1.0/128.0
    f = x*((1 + 0.1*x)**10 - 1)*(1 - x)
    xx = f*y - 0.02*x
    yy = b*y*(1 - y) - f*y - 0.01*y
    x += xx + lap(x,d)
    y += yy + lap(y,d)
    return x, y

def localized_initial_conditions(L):
    x,L2,r = np.zeros((L,L)), L//2, 10
    x[L2-r:L2+r, L2-r:L2+r] = 0.123
    return x, np.random.rand(L,L)

def initial_fig(x, y):
    fig, ax = pl.subplots(1,2,figsize=(11.3,6))
    imx = ax[0].imshow(x, animated=True,vmin=0,cmap='Greys')
    imy = ax[1].imshow(y, animated=True,vmax=1,cmap='Greys')
    ax[0].axis('off')
    ax[1].axis('off')
    ax[0].set_title('x', fontsize=16)
    ax[1].set_title('y', fontsize=16)
    return fig, imx, imy

def fig_update(fid,nn,*args):
    for n in range(nn):
        x, y = rd(*args)
    imx.set_array(x)
    imy.set_array(y)
    imx.set_norm(Normalize(vmin=np.amin(x),vmax=np.amax(x)))
    imy.set_norm(Normalize(vmin=np.amin(y),vmax=np.amax(y)))
    return imx, imy

if __name__ == "__main__":
    L = 256
    x, y = localized_initial_conditions(L)
#    x, y = np.random.rand(L,L), np.random.rand(L,L) 
    fig, imx, imy = initial_fig(x, y)
    a = animation.FuncAnimation(fig, fig_update, fargs=(10, x, y), interval=1, blit=True)
    pl.show()
\end{verbatim}
\end{footnotesize}

\end{document}